# Wide spectrum denoising (WSD) for super-resolution microscopy imaging using compressed sensing and a high-resolution camera

**TAO CHENG**[1,5,*], **DANNI CHEN**[2,3,4,5,6] **& HENG LI**[2,3,4]

[1]*College of Mechanical and Transportation Engineering, Guangxi University of Science and Technology, Liuzhou, Guangxi 545006, China*
[2] *College of Physics and Optoelectronic Engineering, Shenzhen University, Shenzhen, Guangdong 518060, China*
[3] *Key Laboratory of Optoelectronic Devices and Systems of Ministry of Education and Guangdong Province, Shenzhen University, Shenzhen, Guangdong 518060, China*
[4] *Shenzhen Key Laboratory of Micro-Nano Measuring and Imaging in Biomedical Optics of Shenzhen, College of Physics and Optoelectronic Engineering, Shenzhen University, Shenzhen, Guangdong 518060, China*
[5] *Contributed equally to this work.*
[6] *danny@szu.edu.cn*
[*] *ctnp@163.com*

**Abstract:** Because of the lack of effective denoising methods, any form of denoising is seldom performed for super-resolution microscopy, resulting in poor temporal and spatial resolutions. We propose a denoising method for STORM raw images based on compressed sensing and high-resolution cameras. This method overcomes the limitation that the raw pixel size must be approximately equal to the standard deviation of the point spread function. This method can be effectively used to remove random noise such as Poisson and Gaussian noise from very low density to ultra-high density fluorescent molecular distribution scenarios. Therefore, it is a wide spectrum denoising algorithm. Using this method, it was demonstrated that the SNR of a raw image can be increased by approximately 7 dB. Using CVX reconstruction, only 20 frames of the raw image are needed, and the time resolution is 0.86 s. The spatial resolution is also greatly improved.

## 1. Introduction

The noise of a stochastic optical reconstruction microscopy (STORM) raw image acquired by an electron-multiplying charge-coupled device (EMCCD) mainly includes shot noise following a Poisson distribution, readout noise[1, 2] following a Gaussian distribution and the background. Because the readout noise of EMCCD cameras and its variance is small [2-4], STORM simulations sometimes consider only the background and Poisson noise [2-5]. The number of photons detected in one pixel of a raw image follows the Poisson distribution, and the mean is determined by the fluorescent signal in the pixel and the background [2, 4].

Improving the temporal and spatial resolutions of super-resolution microscopy has been the focus of studies on STORM [4, 6, 7]. The existence of noise makes the effective pixel size of the camera approximately equal to the standard deviation (s.d.) of the imaging system's point spread function (PSF) to achieve a better single-molecule localization effect [5, 8]. This tradition is also continued in studies on STORM based on compressed sensing (CS)[4]. If a high-resolution camera (whose effective pixel size is much smaller than the s.d. of the PSF) is used, the number of photons received by each camera pixel will be too small, resulting in increased noise. Therefore, the localization precision will be drastically reduced [5, 9]. Various single-molecule localization algorithms have limited anti-noise abilities and cannot effectively utilize the raw images acquired by high-resolution cameras. CS can be used to acquire and reconstruct raw images of high-density fluorescent molecules and greatly improve the temporal and spatial resolutions of super-resolution microscopy [4, 10-14]. Cheng et al.[15] noted that if a high-resolution camera is used to acquire data, its corresponding measurement matrix has better performance.which is helpful in improving the reconstruction effect. However, if a high-resolution camera is used, the raw image's noise increases and the reconstruction effect simultaneously deteriorates. Therefore, based on CS and high-resolution cameras, the quality of a reconstructed super-resolution image depends on the balance between the performance improvement of the measurement matrix and the increase in the raw image's noise. If the effect of the measurement matrix performance improvement is greater than the adverse effect of the increased noise, the reconstruction effect will increase; otherwise, the opposite will occur [15]. If the noise of the raw image can be effectively denoised, the temporal and spatial resolutions of the CS-based super-resolution microscopy can be further improved. Moreover, the development could further reduce the manufacturing and popularization costs of the related instruments as well as equipment and the difficulty of experimental operation.

There are many excellent high-performance denoising algorithms in the field of microscopy [16] and image processing. These include BM3D (block-matching and 3D filtering) [17] for Gaussian noise, IPVA[18] (iterative Poisson denoising algorithm based on variance stabilization and additive white Gaussian noise denoising)[19] and PABI (denoising Poisson images using the Anscombe variance-stabilizing transformation, BM3D filter, and the exact unbiased inverse of the Anscombe transformation)[20] for Poisson noise, GAV (applying a generalized Anscombe variance-stabilizing transformation to denoising)[21] for Poisson and Gaussian mixed noise, etc.[16, 22, 23]. However, various studies on fluorescent molecular localization in the field of super-resolution microscopy rarely report denoising of the raw image before localization [4, 15, 24]. Single-molecule localization algorithms such as (fluorescence) photoactivated localization microscopy ((f)PALM) perform bandpass filter processing on the raw image before localization [5]. However, the raw image loses a large amount of information, which is not suitable for CS reconstruction and calculation. The raw image of the CS-based STORM [4, 24] is not denoised. The only pre-processing is subtraction of the baseline from the raw image [4, 24]. Hence, the potential of CS cannot be brought into full play. The noise of a STORM raw image is dominated by the Poisson noise, which is mixed with various other types of noises [4, 24]. Although EMCCD performance is steadily improving, readout noise and the like still exist [1, 2].

Super-resolution microscopy has features that distinguish it from microscopy and other imaging environments[25]. Based on CS and high-resolution cameras, an algorithm that is theoretically applicable to various types of random noise was developed for super-resolution microscopy. Because the algorithm is applicable to various types of random noise and the denoising performance is not affected by the distribution density of the fluorescent molecules, we call this algorithm wide spectrum denoising (WSD).

The remainder of this paper is organized as follows. Section 2 introduces the mathematical model of CS based on STORM and the PSF-based measurement matrix. Section 3 presents the derivation of the operator matrix of the PSF-based measurement matrix, proposes the WSD algorithm according to the singular value characteristics of the operator matrix, and provides a theoretical analysis of WSD. Sections 4 and 5 verify the effectiveness of WSD using simulation and through experimental data, respectively. Section 6 describes a method of improving the computing speed through parallel computing. Finally, Section 7 concludes the paper with a brief discussion of related issues.

## 2. Compression sensing and the PSF-based measurement matrix

CS has achieved a great success in STORM-based super-resolution microscopy. If the original image is sparse, CS can reconstruct the super-resolution image with high temporal and spatial resolutions from the STORM raw image [4, 15, 24, 26]. An original image of microtubules and other structures in the cell is essentially sparse. Even if the fluorescent molecules are densely and continuously distributed along the microtubules of a cell, the sparse requirements of CS are still satisfied. However, without denoising pretreatment of the raw image, CS can only reconstruct the raw image with low fluorescence molecular density. The reconstruction potential of CS has not been fully exploited. Equation (1) shows a mathematical model of CS based on STORM [4, 15]. The vectors, $\mathbf{y}$ and $\mathbf{x}$, consist of row-wise or column-wise concatenations of the raw image $\mathbf{Y}$ (i.e., the camera image) and the super-resolution image $\mathbf{X}$ (i.e., the pixelated original image) respectively. The measurement matrix $\mathbf{A}$ is determined by the PSF of the imaging system. Therefore, the measurement matrix $\mathbf{A}$ is a PSF-based measurement matrix. For the convenience of description, the PSF-based measurement matrix is hereinafter referred to as the measurement matrix for short, too. The acquired raw image corresponds to the $i^{th}$ column of $\mathbf{A}$ if only one molecule emits fluoroscopic photons at the position index $i$ of $\mathbf{x}$ [4]. Using the CVX software package [4, 27], the super-resolution image can be solved by equation (1). The number of non-zero elements in $\mathbf{x}$ is called sparsity in CS, and it is represented by K.

$$\min \|\mathbf{c}^T \mathbf{x}\|_1 \quad s.t. \quad \|\mathbf{A}\mathbf{x} - \mathbf{y}\|_2 \leq \varepsilon \sqrt{\sum \mathbf{y}_j} \quad and \quad \mathbf{x}_i \geq 0, \tag{1}$$

where $\mathbf{c}, \mathbf{x} \in \mathbf{R}^N$; $\mathbf{y} \in \mathbf{R}^M$; $\mathbf{A} \in \mathbf{R}^{M \times N}$; and $M<N$, with $N$ and $M$ being natural numbers. $\mathbf{c}$ and $\mathbf{x}$ are column vectors containing $N$ elements. $\mathbf{c}^T$ is a row vector and the transpose matrix of $\mathbf{c}$. $\mathbf{y}$ is a column vector containing $M$ elements. $\mathbf{A}$ is a matrix of size $M \times N$. The objective function is $\min \|\mathbf{c}^T \mathbf{x}\|_1$ to obtain the minimum value of $\|\mathbf{c}^T \mathbf{x}\|_1$. $\mathbf{c}^T \mathbf{x}$ is the dot product of two vectors, $\mathbf{c}$ and $\mathbf{x}$. The constraint functions are $\|\mathbf{A}\mathbf{x} - \mathbf{y}\|_2 \leq \varepsilon \sqrt{\sum \mathbf{y}_j}$ and $\mathbf{x}_i \geq 0$. The abbreviation 's.t.' stands for 'subject to' which is used to derive constraint functions in the CS. $\mathbf{y}_j$ is the $j^{th}$ element of $\mathbf{y}$. $\mathbf{x}_i$ is the $i^{th}$ element of $\mathbf{x}$. $\|\cdot\|_1$ is the $l_1$-norm (the sum of the element magnitudes) of a vector (e.g., $\|\mathbf{x}\|_1 = |\mathbf{x}_1| + |\mathbf{x}_2| + \cdots + |\mathbf{x}_N|$). $\|\cdot\|_2$ is the $l_2$-norm (e.g., $\|\mathbf{x}\|_2 = \sqrt{|\mathbf{x}_1|^2 + |\mathbf{x}_2|^2 + \cdots + |\mathbf{x}_N|^2}$) of a vector. The weight vector, $\mathbf{c}$, is introduced to account for the differing contributions to the raw image from one single fluorescent molecule at different locations. The value of the $i^{th}$ element of $\mathbf{c}$ is equal to the summation of the $i^{th}$ column of the measurement matrix $\mathbf{A}$. $\varepsilon$ is a constant. The value of $\varepsilon$ was 2.1 [4] for actual experimental data in a previous study, and this value was also used in the experiment in this study. One additional element in $\mathbf{x}$ is introduced if the uniform background of the image is not zero. The corresponding element in $\mathbf{c}$ is set to 0, and all elements in the corresponding column of $\mathbf{A}$ are set to 1 [4, 15].

For convenient description, $\mathbf{A}$ with one column whose elements are all 1 is denoted by $\mathbf{\Phi}$, where $\mathbf{\Phi} \in \mathbf{R}^{M \times (N+1)}$. The measurement matrix $\mathbf{A}$ without a column whose elements are all 1 is still denoted by $\mathbf{A}$, where $\mathbf{A} \in \mathbf{R}^{M \times N}$. A STORM raw image acquired by a high-resolution camera corresponded to the measurement matrix $\mathbf{A}$ whose size is 196 × 4096 in this study. For distinction from the pixels of the raw image, the pixels of the super-resolution image are referred to as grids. The super-resolution image's grid is 1/4 of the pixel size of the raw image. The size of the measurement matrix $\mathbf{A}$, corresponding to the binned raw image, is 49 × 4096.

In the CS model shown in equation (1), the uniform background should be treated as part of the signal rather than the noise. If only a uniform background exists and no other noise is present, $\mathbf{\Phi}\mathbf{x} = \mathbf{y}$. In all the signal-to-noise ratio (SNR) calculations in this study, the uniform background was considered to be part of the signal.

## 3. Wide spectrum denoising

### 3.1 Operator matrix of the PSF-based measurement matrix and high-dimensional ellipsoid projection

The row-column ratio of the measurement matrix **A** based on STORM is only 196:4096, even with a high-resolution camera. The columns are highly similar [15] with a maximum column correlation coefficient of 0.999224. The measurement matrix **A** is a highly flat autocorrelation matrix. Therefore, it has unique properties.

If the measurement matrix **A** is operated by orthogonalization and normalization using MATLAB, the measurement matrix $\mathbf{A}_O$ can be obtained. Row orthogonal normalization is an operation in the matrix theory. It makes the rows of the matrix completely orthogonal. Moreover, it makes the $l_2$-norm of each row of the matrix equal to 1. Through $\mathbf{A}_O$ and **A**, the operator matrix **T** which is equivalent to the row orthogonal normalization operation can be obtained, where $\mathbf{T} = \mathbf{A}_O \mathbf{A}^T (\mathbf{A}\mathbf{A}^T)^{-1}$. Therefore, $\mathbf{y} = \mathbf{A}\mathbf{x}$ can be equivalently converted into $\mathbf{T}\mathbf{y} = \mathbf{T}\mathbf{A}\mathbf{x}$. If a singular value decomposition is applied to **T**, then $\mathbf{T} = \mathbf{U}\mathbf{S}\mathbf{V}^T$, where **S** is a diagonal sparse matrix that is composed of singular values of **T**; $\mathbf{V}^T$ is a transpose matrix of **V**, and **T**, **U**, **S**, and **V** are square matrices. Therefore, $\mathbf{T}\mathbf{y} = \mathbf{T}\mathbf{A}\mathbf{x}$ can be equivalently converted into $\mathbf{S}\mathbf{V}^T\mathbf{y} = \mathbf{S}\mathbf{V}^T\mathbf{A}\mathbf{x}$. According to the geometrical and physical meanings of the square matrices **S** and **V**, **V** can change the direction of a vector, and **S** can change the magnitude (i.e., $l_2$-norm) of a vector [28, 29]. **S** can be regarded as a high-dimensional ellipsoid projection operator with the singular values of **T** as the axis of the high-dimensional ellipsoid. The 196 singular values of **T** is shown in Fig. 1. The ratio of the long axis to the short axis of the high-dimensional ellipsoid is $5.792 \times 10^{12}$.

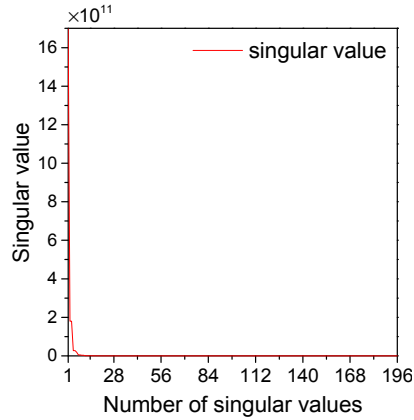

**Fig. 1.** Analysis of singular values of the operator matrix **T**. The maximum and minimum singular values of **T** are $1.697 \times 10^{12}$ and 0.293 respectively.

### 3.2 Wide spectrum denoising algorithm

The detailed WSD algorithm of a STORM raw image is shown in Algorithm 1. We will clearly explain the principle of WSD and the denoising process by Fig. 2.

**Algorithm 1:** Wide Spectrum Denoising (WSD) Algorithm for Super-Resolution Microscopy Imaging Using Compressed Sensing and a High-Resolution Camera

**Input:**

1) $\mathbf{A}$, a PSF-based measurement matrix; $\mathbf{Y}_{raw}$, a STORM raw image; star=0.9.

2) tail=0.95 (see Section 3.3 for more explanation of star and tail); M, the number of rows in $\mathbf{A}$ and the number of pixels of $\mathbf{Y}_{raw}$.

**Denoising:**

1) The vector, $\mathbf{y}_{raw}$, consists of column-wise (or row-wise) concatenations of $\mathbf{Y}_{raw}$.

2) $\mathbf{A}_O$, row orthogonal normalization of $\mathbf{A}$.

3) $\mathbf{T} = \mathbf{A}_O \mathbf{A}^T (\mathbf{A}\mathbf{A}^T)^{-1}$, where $\mathbf{A}_O = \mathbf{T}\mathbf{A}$, and $\mathbf{U}\mathbf{S}\mathbf{V}^T$ is the singular value decomposition of the operator matrix $\mathbf{T}$, where $\mathbf{T} = \mathbf{U}\mathbf{S}\mathbf{V}^T$.

4) $\mathbf{z}_{raw} = \mathbf{S}\mathbf{V}^T \mathbf{y}_{raw}$.

5) $i_{star}$ =[M×star], $i_{tail}$ =[M×tail] (see Section 3.3 for more explanation of the '[ ]').

6) Threshold value cri: the largest magnitude entries corresponding to the indices from $i_{star}$ to $i_{tail}$ in the absolute values of the vector $\mathbf{z}_{raw}$ (see Section 3.3 for more explanation of the cri).

7) $\mathbf{I}_{upper}$ refers to indices corresponding to $\mathbf{z}_{raw}$'s magnitude elements greater than cri. $\mathbf{z}_{raw}$'s magnitude elements whose indices belong to $\mathbf{I}_{upper}$ are set to cri. The new $\mathbf{z}_{raw}$ is represented by $\mathbf{z}_{WSD}$ (see Section 3.3 for more).

8) $\mathbf{y}_{WSD} = \mathbf{T}^{-1}\mathbf{U}\mathbf{z}_{WSD}$.

**Output:**

1) The matrix, $\mathbf{Y}_{WSD}$, consists of the vectors $\mathbf{y}_{WSD}$ by reshaping (reshaping can be achieved by the reshape function in MATLAB).

Fig. 2 presents the denoising principle of WSD and the denoising process analysis of the raw image ($\mathbf{Y}_{raw}$, $\mathbf{Y}_{raw} \in \mathbf{R}^{\sqrt{M} \times \sqrt{M}}$, $M = 196$, as shown in Fig. 2e), containing four fluorescent molecules. The simulation was conducted for an average photon number of 3,000 per molecule and a background of 16 photons per pixel with Poisson noise. For Fig. 2, $\mathbf{y}_{raw}$, where $\mathbf{y}_{raw} \in \mathbf{R}^M$, is a vector that consists of column-wise concatenations of $\mathbf{Y}_{raw}$ of a raw image, as do $\mathbf{y}_{ini}$ and $\mathbf{y}_{WSD}$. For the convenience of description, the $\mathbf{Y}_{raw}$ and $\mathbf{y}_{raw}$ are hereinafter referred to as the raw image. $\mathbf{Y}_{WSD}$ is the denoised $\mathbf{Y}_{raw}$ by WSD. $\mathbf{Y}_{ini}$ is the noiseless raw image ($\mathbf{Y}_{raw}$) with a uniform background, as shown in Fig. 2e.

Fig. 2a shows a comparative analysis diagram of $\mathbf{y}_{raw}$ and $\mathbf{y}_{ini}$. The curves of $\mathbf{y}_{raw}$ and $\mathbf{y}_{ini}$ are inconsistent because the former contains noise and the latter does not. For example, the crest difference in the dotted ellipse is 24. The SNR of $\mathbf{y}_{raw}$ is only 20.35 dB. The crests are located approximately at the centers of $\mathbf{Y}_{raw}$ and $\mathbf{Y}_{ini}$.

The two curves ($\mathbf{z}_{raw}$ and $\mathbf{z}_{ini}$) shown in Fig. 2c were obtained by $\mathbf{z}_{raw} = \mathbf{S}\mathbf{V}^T\mathbf{y}_{raw}$ and $\mathbf{z}_{ini} = \mathbf{S}\mathbf{V}^T\mathbf{y}_{ini}$ from Fig. 2a. $\mathbf{U}\mathbf{S}\mathbf{V}^T$ is a singular value decomposition of the operator matrix $\mathbf{T}$. The curves $\mathbf{z}_{ini}$ in Figs. 2c and 2d are the same. The range of the y-axis in Fig. 2d is only from -20,000 to 10,000. Meanwhile, the range of the y-axis in Fig. 2c is from $-2 \times 10^{12}$ to $1 \times 10^{12}$, which is $10^8$ times that of Fig. 2d. Therefore, the curve $\mathbf{z}_{ini}$ appears as a straight line in Fig. 2c and as a wave line in Fig. 2d. The curve $\mathbf{z}_{ini}$ (which fluctuates up and down along a line with a y-axis coordinate of 0) in Fig. 2d is very stable except for a few crest and troughs that are marked 1, 2, 3, 4, 5, and 6. Through numerous experiments, we determined the amplitude of the stable fluctuation of the curve $\mathbf{z}_{ini}$ in Fig. 2d. The amplitude is the threshold value cri of Algorithm 1 (see Section 3.3 for details).

The elements greater than cri in $\mathbf{z}_{raw}$ contain noise. We replaced these elements with cri. Then $\mathbf{z}_{WSD}$ (i.e. the new $\mathbf{z}_{raw}$) in Fig. 2d is obtained. As shown in Fig. 2d, except for a few crest and troughs which are marked with 1, 2, 3, 4, 5, and 6, the amplitudes of the curves $\mathbf{z}_{WSD}$ and $\mathbf{z}_{ini}$ are very close.

According to $\mathbf{y}_{WSD} = \mathbf{T}^{-1}\mathbf{U}\mathbf{z}_{WSD}$ of Algorithm 1, $\mathbf{y}_{WSD}$ of Fig. 2b can be obtained from $\mathbf{z}_{WSD}$ of Fig. 2d. The crest difference in the dotted ellipse is only 6. The curves of $\mathbf{y}_{WSD}$ and $\mathbf{y}_{ini}$ shown in Fig. 2b agree well because the former is denoised by WSD. Moreover, they almost completely coincide. The SNR of $\mathbf{y}_{WSD}$ is 29.07 dB which is 8.72 dB higher than that of $\mathbf{y}_{raw}$ in Fig. 2a.

Fig. 2e shows a noise-free raw image and raw images before and after denoising corresponding to Figs. 2a and b and the corresponding binned images. Comparison of $\mathbf{Y}_{raw}$, $\mathbf{Y}_{WSD}$ and $\mathbf{Y}_{ini}$, reveals that the denoising effect of the raw image is obvious, and the SNR improves by 8.72 dB. Thus, the noise component of $\mathbf{Y}_{raw}$ has been substantially eliminated. The denoised raw image ($\mathbf{Y}_{WSD}$) is very similar to $\mathbf{Y}_{ini}$. Comparison of the binned $\mathbf{Y}_{raw}$, $\mathbf{Y}_{WSD}$ and $\mathbf{Y}_{ini}$ (i.e., $\mathbf{Y}_{raw\_bin}$, $\mathbf{Y}_{WSD\_bin}$ and $\mathbf{Y}_{ini\_bin}$) shows that the denoising effect is not obvious, and the SNR increases by 3.34 dB. For different sparsities and noise levels, similar patterns were obtained. Fig. 2e shows the simulated denoising results after taking the logarithms for visualization. This approach (a frequently used data processing method in vibration mechanics [30] and digital image processing [31]) enables easy checking and observation of data. Moreover, it preserves the relative scale of the data.

### *3.3 Selection of star, tail, and threshold value cri*

The inset in Fig. 2c shows a partial enlargement of the right ends of the curves ($\mathbf{z}_{raw}$ and $\mathbf{z}_{ini}$) marked by the dotted ellipse of Fig. 2c. The range of the y-axis in the inset is from -750 to 750. As shown in the inset, the right ends of the curves $\mathbf{z}_{raw}$ and $\mathbf{z}_{ini}$ with the x-axis (i.e. abscissa) from 176 to 196 are almost the same. In the rest of the curves, the amplitude of the curve $\mathbf{z}_{raw}$ is far larger than that of $\mathbf{z}_{ini}$. The tendency of this amplitude of $\mathbf{z}_{raw}$ to increase gradually becomes stronger as it approaches the coordinate origin. The range of the y-axis of the curves ($\mathbf{z}_{raw}$) in Fig. 2c is from $-2\times10^{12}$ -$1\times10^{12}$. The range of the y-axis of the curves ($\mathbf{z}_{ini}$) in Fig. 2d is from -20000-10000. For high-density fluorescent molecules, the right ends of the curves $\mathbf{z}_{raw}$ and $\mathbf{z}_{ini}$ with the x-axis coordinate of 196 are second only to the crest and troughs that are marked with 1, 2, 3, 4, 5, and 6 in Fig. 2d. We performed numerous experiments and verified that this phenomenon is stable. Therefore, we use the maximum absolute value of the curve ($\mathbf{z}_{raw}$) in the x-axis from 176 to 186 as the threshold value of cri where 176 is the nearest integer less than or equal to 196 × 0.9, and 186 is the nearest integer less than or equal to 196 × 0.95. The symbol '[ ]' is a rounding symbol, i.e., [x] represents a nearest integer less than or equal to x in Algorithm 1. Therefore, star=0.9 and tail=0.95, as obtained empirically through many simulation experiments. If more experiments were performed, better star and tail values could be found.

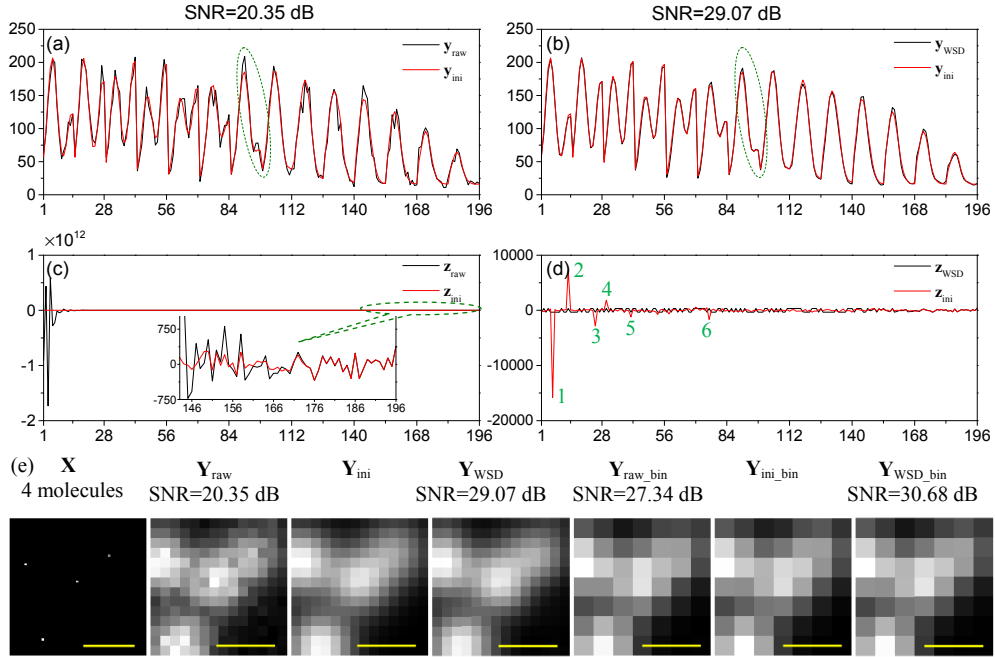

**Fig. 2.** Denoising analysis of simulated STORM raw image using WSD. (a) - (d) Denoising process analysis. The y-axis represents the values of the elements of the vectors $\mathbf{y}_{ini}$, $\mathbf{y}_{WSD}$, $\mathbf{y}_{raw}$, $\mathbf{z}_{raw}$, $\mathbf{z}_{ini}$, and $\mathbf{z}_{WSD}$. The range of the y-axis in (c) is from $-2\times 10^{12}$ to $1\times 10^{12}$. The inset in (c) is an enlarged view from the end area marked by the dotted ellipse and has a y-axis range of -750–750. The range of the y-axis in (d) is from -20000 to 10000. The serial numbers 1, 2, 3, 4, 5, and 6 in (d) are the numbers of the six largest crests and troughs of the curve $\mathbf{z}_{ini}$. The x-axis is labeled with numbers from 1 to 196. The number of elements in $\mathbf{y}_{ini}$, $\mathbf{y}_{WSD}$, $\mathbf{y}_{raw}$, $\mathbf{z}_{ini}$, $\mathbf{z}_{WSD}$, and $\mathbf{z}_{raw}$ is 196. (e) The noiseless raw image and raw images before and after denoising by WSD and the corresponding binned raw images. The $\mathbf{X}$ is the original image with the known positions of 4 molecules. Scale bars, 274 nm.

*3.4 Theoretical analysis of WSD*

Regardless of the type of noise, it is approximately orthogonal to the signal as long as it is random noise. As shown in Fig. 2c, the noise-free raw image was projected onto the short-axis region of the high-dimensional ellipsoid of $\mathbf{S}$. Thereafter, the noise that was approximately orthogonal to this image was projected onto the long-axis region of the high-dimensional ellipsoid of $\mathbf{S}$, resulting in a sharp amplification of the noise. S is a sparse matrix whose elements are zero except for the diagonal elements. In the matrix theory, such a matrix is referred to as a high-dimensional ellipsoid with diagonal elements as its long- and short-axes. In matrix theory, $\mathbf{z}_{raw} = \mathbf{S}\mathbf{q}_{raw}$ is called, projecting $\mathbf{q}_{raw}$ onto the high-dimensional ellipsoid $\mathbf{S}$, where $\mathbf{q}_{raw} = \mathbf{V}^T\mathbf{y}_{raw}$. Therefore, the denoising performance of WSD is independent of the random noise type. Regardless of the molecular density in the raw image, the approximately orthogonal relationship between the noise in the raw image and the noise-free raw image is not changed. This indicates that the denoising performance of the WSD is independent of the molecular density.

Because the measurement matrix $\mathbf{A}$ is a highly flat autocorrelation matrix, the calculation accuracy of MATLAB is insufficient. When solving $\mathbf{T}$, there could be some MATLAB issues ('Warning: Matrix is close to singular or badly scaled. Results may be inaccurate. RCOND =

4.517403e-20.'). To obtain effective calculation results, we used the Multiprecision Computing Toolbox to improve the calculation accuracy to 40 decimals.

## 4. Simulation data analysis of the WSD

Regardless of the sparsity (molecular density), combination, and level of noise, the denoising effect of the WSD is superior to those of the existing mature denoising algorithms, including BM3D, IPVA, PABI and GAV. The denoising performance of the WSD has nothing to do with the molecular density and type of random noise. WSD can always achieve a denoising effect of approximately 7 dB. Such stability is not achievable using BM3D, IPVA, PABI and GAV. The SNR of the denoised raw image by the WSD is larger than that of the binned raw image, as shown in Fig. 3. Thus, the use of a high-resolution camera can provide better image quality based on WSD than those currently achievable using low-resolution cameras. Therefore, the limitation that the effective pixel size of the camera must be equal to the PSF's s.d. is broken through [5].

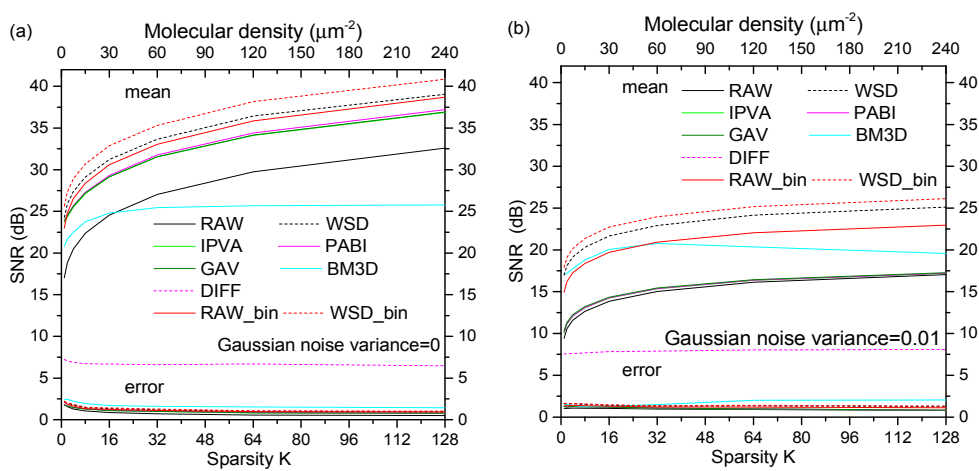

**Fig. 3.** Comparison and analysis of the denoising efficiency of STORM raw image using WSD and other denoising methods. IPVA, PABI, GAV, and BM3D are four different denoising algorithms for raw images (RAW). The WSD_bin indicates the binned raw image after denoising using WSD. RAW represents the raw image. RAW_bin indicates the binned raw image. The 'mean' is the mean SNR. The 'error' is the s.d. of the SNR. DIFF is the difference between the means of the WSD and RAW. The x-axis indicates both molecular density and signal sparsity. The y-axis represents the SNR. (a) The variance of the Gaussian noise is 0. That is, there is no Gaussian noise, only Poisson noise. (b) The variance of the Gaussian noise is 0.01. That is, there are Gaussian noise and Poisson noise.

*4.1 Simulation data*

To evaluate the denoising performance of WSD, we generated simulated images with known real molecular positions. The simulation randomly placed K molecules in a 64×64 grid (0.731 μm × 0.731 μm) region. K represents the sparsity in CS and was equal to 1, 2, 4, 8, 16, 32, 64 or 128, with corresponding molecular densities of 1.869, 3.738, 7.477, 14.954, 29.907, 59.815, 119.629, and 239.258 $\mu m^{-2}$, respectively. The grid size was 11.43 nm. The effective pixel size of the raw image, 45.714 nm, matched the pixel size of our microscope setup and was roughly equal to half of the s.d. of the PSF of our setup. The size of the raw image was 14×14 pixels in the middle of the 64×64 grid super-resolution image. At this point, the size of the corresponding measurement matrix **A** was 196×4096. The size of the binned raw image was 7×7 pixels in the middle of the 64×64 grid super-resolution image. The PSF of the microscope was a Gaussian function. In a STORM simulation experiment, the number of photons detected from each molecule followed a log-normal distribution to approximate the experimentally measured single-molecule photon number distribution [4].

In our simulations, we set the peak of the log-normal photon number distribution as 3,000 photons (with an s.d. of 1,700 photons). The uniform background of the raw images was 16 photons per pixel. The uniform background of the binned raw image was 64 photons per pixel. Poisson and Gaussian noises (Gaussian noise variances of 0.01) were also added. For each K at different noise levels and different denoising methods, the simulation was repeated 500 times as shown in Fig. 3. Thereafter, the mean and s.d. of the SNR of the various denoising algorithms were calculated. Fig. 2 is one simulation of Fig. 3a.

*4.2 Simulation results*

For the raw images with only uniform backgrounds and Poisson noise as well as molecular sparsity K of 1-128, we analyzed the denoising performance of the WSD and other denoising algorithms (Fig. 3a). The upper part of DIFF is the mean of the SNRs of various denoising algorithms, and the lower part is the s.d. of the SNRs of various denoising algorithms. The maximum value of each s.d. curve does not exceed 2.5 dB. Thus, the performances of the various algorithms are stable. The near-horizontal curve DIFF is the difference between the means before and after denoising the raw image by the WSD. As can be seen from DIFF, the denoising ability of the WSD is very stable and does not change with the molecular density. The SNR improves by approximately 7 dB. According to the mean curves, IPVA and PABI achieved good denoising effects for the Poisson noise, and the SNR improved by approximately 5 dB. GAV also achieved a denoising effect similar to those of IPVA and PABI for the Poisson and Gaussian mixed noise. The three curves almost completely coincide. The BM3D for the Gaussian noise has a small denoising effect when the sparsity K < 20. This algorithm is completely ineffective when K > 20. The WSD completely surpasses the denoising effects of IPVA, PABI, and GAV, and also achieves a better SNR than that of the binned raw image (RAW_bin). The SNR of the binned denoised raw image (WSD_bin) is better than that of the binned raw image (RAW_bin), and the SNR improves by approximately 2 dB. The SNR of the binned denoised raw image (WSD_bin) is also better than that of the denoised raw image (WSD).

The performance analysis curve of each denoising algorithm is shown in Fig. 3b after further adding Gaussian noise with a variance of 0.01 based on Fig. 3a. The Gaussian noise with a variance of 0.01 is already large in general image processing. The maximum value of each s.d. curve is approximately 2 dB, which indicates that the performances of the various algorithms are stable. It can be seen from DIFF that the denoising ability of the WSD is very stable and does not change with molecular density. The SNR improves by approximately 7.5 dB or more. According to the mean curves, BM3D has a good denoising effect in the interval in which the sparsity is less than 30. The SNR after denoising exceeds that of the binned raw image (RAW_bin). However, the denoising ability of BM3D varies with the sparsity. Although the SNR is still higher than that of the raw image (RAW) in the interval in which the sparsity is greater than 30, it has a downward trend. IPVA and PABI are suitable only for the Poisson noise. In the case of Gaussian noise, the SNR curves of IPVA and the PABI are still above that of the raw image (RAW). However, these algorithms yield minimal SNR improvement of less than 1 dB and thus are almost ineffective. Although GAV is theoretically applicable to scenarios involving mixtures of Poisson and Gaussian noise, GAV is similar to IPVA and PABI in the context of super-resolution microscopy. The three curves are almost completely coincidental. The WSD curve is completely above that of BM3D and RAW_bin. The SNR of WSD_bin is approximately 3 dB higher than that of RAW_bin. The SNR of the WSD_bin is also better than that of the WSD.

## 5. Experimental data analysis of the WSD

An actual denoising experiment of WSD proved that a complex cell microtubule structure could be reconstructed using 20 frames of the raw image and the CVX algorithm (Fig. 4). The raw images were recorded with an EMCCD camera at a frame rate of 23.218 Hz. The time resolution of the super-resolution image sequence reaches approximately 0.8614 s (20 frames). Fig. 4 shows the results before and after denoising of the real raw images and the corresponding reconstruction results by the CVX after taking the logarithms for visualization.

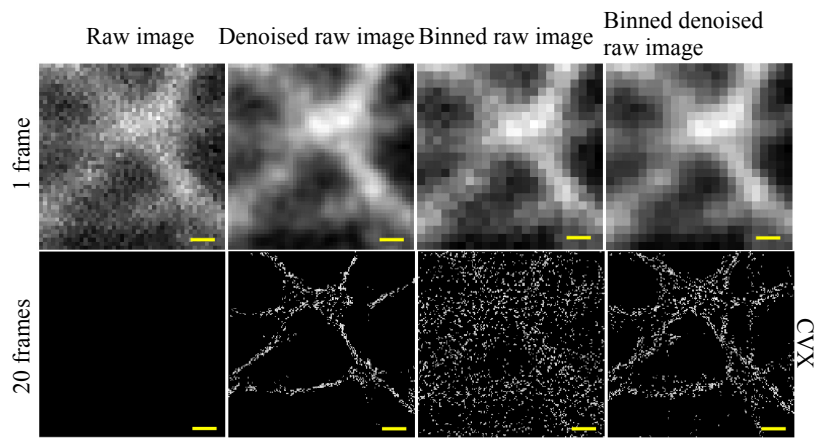

**Fig. 4.** Denoising and reconstruction of experimental STORM raw images using the WSD and CVX. The leftmost panel of the upper row (one fluorescence raw image frame from 20 frames captured during STORM data acquisition) shows the high density of the activated fluorophores. The other three panels are the denoised raw image using the WSD, the binned raw image and the binned denoised raw image by WSD respectively. Lower row: super-resolution images reconstructed from 20 frames of raw images corresponding to the upper row by CVX. Scale bars, 274 nm.

*5.1 Sample preparation*

HeLa cells were grown on the coverslips of approximately 80% confluence, and fixed with ice-cold of 4% formaldehyde/PBS (freshly prepared from paraformaldehyde) for 10 min. Thereafter, they were permeabilized and blocked at room temperature with 0.3% Triton-X 100 for 20 min and 2% bovine serum albumin for 1 h. Staining of tubulin was performed with a mouse monoclonal anti–α-tubulin antibody and 2% bovine serum albumin (1:1000, 60 min incubation at room temperature; Sigma), a secondary donkey anti-mouse antibody and 2% bovine serum albumin (1:3000, 60 min incubation at room temperature).

*5.2 Optical setup and imaging*

The optical setup was constructed based on an inverted optical microscope (IX-71, Olympus) with a 100× oil immersion objective lens (UplanSApo, N.A. 1.40, Olympus). A 641-nm laser (CUBE 640-100C; Coherent) was used for fluorescence excitation of Alexa-647. A highly inclined and laminated optical sheet (HILO) configuration was used to diminish the fluorescence background from the non-focal plane. In the imaging optical path, the fluorescence was collected with the same objective. Thereafter, a 45° dichroic beam-splitter (ZT647rdc, Chroma) was used to separate the fluorescence from the laser. Following this, a bandpass fluorescence filter (FF01-676/37, Semrock) was used to filter the fluorescence. An additional relay system with a 3.5-fold image magnification was inserted between the original image plane of the microscope and the detector, and was composed of two lenses whose focal lengths were 50 cm and 175 cm respectively. The fluorescent signals were finally acquired using an EMCCD (DU-897U-CV0, Andor). The exposure time for each raw image was set to 30 ms. Twenty frames were collected for the reconstruction. During the whole image processing, an anti-drift device was used to keep the drift of the focal plane within 10 nm.

*5.3 Experimental results*

To avoid edge effects and speed up the calculation, we extracted partially overlapping small patches (we subdivided the experimental STORM raw image into smaller patches) from the experimental STORM raw image of each frame and denoised each patch using the WSD algorithm. Thereafter, the denoised raw images were merged patch by patch after the overlapping portions of the patches were cut off. For example, a 14 × 14 pixel patch was extracted from the experimental STORM raw image. Only the middle 10 × 10 pixel area was reserved after denoising by WSD (Fig. 4). The CVX was also similar[4] (Fig. 4).

Fig. 4 shows an experimental raw image with a high-density molecular distribution before and after denoising by the WSD and binning. Comparison of the images reveals that the denoised raw image is smoother than the raw image, and the denoising effect is obvious. The simulation analysis shown in Fig. 3 contains only a uniform background as well as Poisson and Gaussian noise. The experimental STORM raw image contains more complex and variable noise types. As shown by the denoising effect and the reconstruction effect in Fig. 4, WSD can adapt to the complex application environment of an experimental STORM raw image.

Because the raw image noise is large before denoising, the CVX reconstruction result of the 20 frames is NaN in MATLAB. Correspondingly, the black image appearing in Fig. 4 indicates that the reconstruction failed. Comparison of the reconstruction results of the raw image before and after denoising in Fig. 4 reveals that the spatial resolution is obviously improved and the cell microtubule structure is clearer.

After denoising, the CVX algorithm can converge faster, and the number of iterations is reduced, increasing the computational speed. As shown in Fig. 4, the CVX calculation time of one frame of the raw image is 119 min and 101 min before and after denoising, respectively. The corresponding CVX calculation times after binning were 12 min and 10 min, which are reduced by 18 min and 2 min, respectively. Single-frame denoising takes 2 min, while the high-precision calculation of solving the operator matrix takes less than 1 min.

## 6. Implementation of the WSD

Current CPUs are multi-core. Multi-core parallel computing can make full use of computing power and save computing time. In the simulation experiments, we used the parfor loop statement of MATLAB parallel computing to realize the denoising calculation of raw images. In the real experiments, based on the parfor loop statement of MATLAB, the parallel denoising and CVX reconstruction calculation of all frames was realized. The computation time of our WSD algorithm depends on the size of the raw image, and it has nothing to do with the number of molecules per frame.

## 7. Conclusion

Various types of random noise and signals naturally have approximate orthogonality. The theoretical basis of the WSD is to exploit this approximate orthogonality. The results proves that the WSD can allow STORM to monitor living cell life processes with sub-second resolution. The spatial resolution is also greatly improved. The limitation that the raw pixel size must be approximately equal to the PSF's s.d. can be broken by the WSD [5]. The experiments showed that the WSD can be used for molecular distribution scenarios from very low density to ultra-high density fluorescence with mixed noise, and can stably increase the SNR of a raw image by approximately 7 dB. The WSD reduces the calculation time of the CVX.

If the WSD algorithm is deeply studied, it is expected to further improve the denoising ability of the WSD and to enhance the reconstruction density as well as the localization precision of the CS to achieve greater improvement in the temporal and spatial resolutions of STORM. Hence, this method provides new possibilities and tools for research and development in super-resolution microscopy.

The WSD has broad application prospects in other fields of super-resolution microscopy such as (f)PALM, stimulated emission depletion (STED), reversible saturable optical linear fluorescence transitions (RESOLFT), super-resolution optical fluctuation imaging (SOFI), and SPoD/ExPAN [15].

## Funding

National Natural Science Foundation of China (81660296, 41461082, 11774242 and 61335001); China Postdoctoral Science Foundation (2016M592525); Specially Funded Program on National Key Scientific Instruments and Equipment Development (2012YQ15009203); and Shenzhen Science and Technology Planning Project (JCYJ20170818142804605).

## Acknowledgements

We thank Tianhe-2 supercomputer for generously providing the high performance computing service.

## Disclosures

The authors declare no competing financial interests.